\documentclass[a4paper,11pt]{article}
\pdfoutput=1 

\usepackage{jinstpub} 

\title{A monitoring and control test bench for assessing the upgraded low voltage power supplies for the ATLAS Tile Calorimeter Phase-II Upgrade front-end electronics}

\author[a]{E. Nkadimeng,\note{edward.khomotso.nkadimeng@cern.ch}}
\author[a]{R. Mckenzie,\note{ryan.peter.mckenzie@cern.ch}}
\author[b]{S. Moayedi,\note{seyedali.moayedi@cern.ch}}
\author[c]{S. Nemecek,\note{Stanislav.Nemecek@cern.ch}}
\author[b]{H. Hadavand,\note{Haleh.Hadavand@cern.ch}}
\author[a,d]{B. Mellado,\note{Bruce.Mellado.Garcia@cern.ch}}
\author[a]{R. van Rensburg\note{roger.mc.lennon.van.rensburg@cern.ch}}

\affiliation[a]{ School of Physics and Institute for Collider Particle Physics, University of the Witwatersrand, Johannesburg, Wits 2050, South Africa}
\affiliation[b]{University of Texas at Arlington, Department of Physics, 502 Yates Street, Arlington, TX 76019, United States of America}
\affiliation[c]{Czech Academy of Sciences (CZ), Czech Republic}
\affiliation[d]{ iThemba LABS, National Research Foundation, PO Box 722, Somerset West 7129, South Africa}

\emailAdd{edward.khomotso.nkadimeng@cern.ch}
\abstract{The ATLAS detector is set to undergo a substantial upgrade termed the "Phase-II" upgrade during the Long-Shutdown in preparation for the start of operation of the
High Luminosity Large Hadron Collider (HL-LHC). This paper describes the development and implementation of the Phase-II upgrade low voltage power supply transformer-coupled buck converter (Brick) test bench. 
A large scale production of approximately 2048 finger low voltage power supplies, with an identical output voltage, is set to be undertaken in the year 2022. This production will culminate in the complete replacement of the LVPS Brick V7.5.0 currently installed within the Tile Calorimeter. Due to the location of the Bricks within the inner barrel of the TileCal their reliability is of the utmost importance. Therefore, the implementation of performance screening will be implemented and will involve the testing and certification of all Bricks produced. A custom test bench has been developed in order to facilitate the above-mentioned testing. The test bench described herein quantifies and logs the performance of an upgrade Brick allowing for comparison with predefined performance metrics before being approved and subsequently installed on-detector. 

}

\arxivnumber{1234.56789} 

\begin{document}
\maketitle

\section{ATLAS experiment for the HL-LHC}
\begin{figure}[t]
\includegraphics[width=11.5cm]{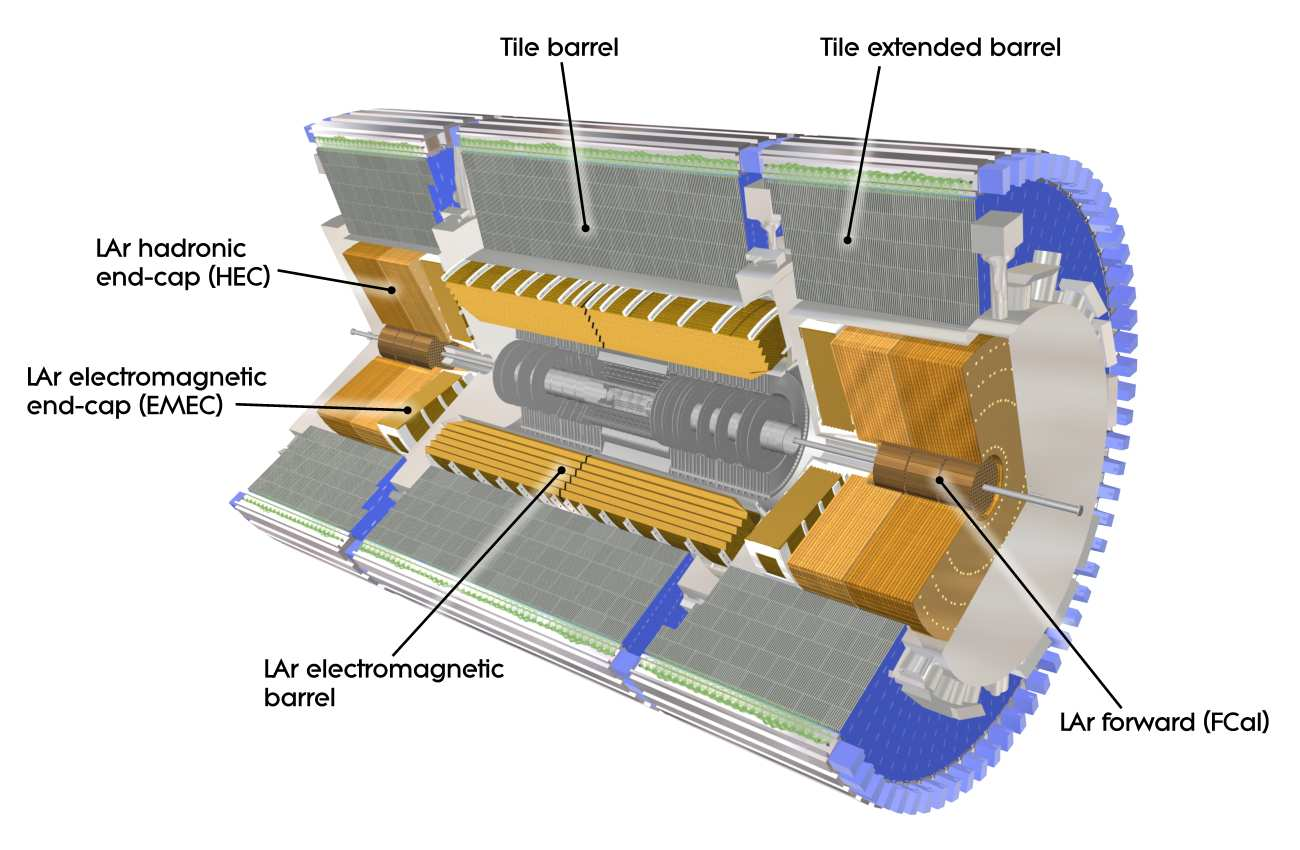}
\centering
\caption{ATLAS inner-barrel~\cite{ref1}.}
\label{fig1}
\end{figure}

The Large Hadron Collider (LHC) accelerator~\cite{ref1}, is scheduled to undergo a major upgrade, in preparation for the High Luminosity LHC (HL-LHC) operation in 2029. The HL-LHC's peak luminosity is expected to increase significantly beyond that of the initial design value. The increase in integrated luminosity, will correspond to an average of 200 simultaneous proton-proton interactions per bunch crossing. The ATLAS detector~\cite{ref1,ref2} surrounds the interaction region at Point 1 of the LHC and is one of the two general-purpose particle detectors.  The purpose of ATLAS is to identify particles produced by protons colliding or heavy ions. Its goal is to measure the properties of these particles. The Tile Calorimeter, as seen in Figure~\ref{fig1}, covers the central region of the ATLAS detector and is a sampling calorimeter composed of scintillating tiles as the active material and steel plates as the absorber. It is divided into three cylinders, namely the Long-Barrel (LB) which covers the region |$\eta$| < 1.0, while the Extended-Barrels (EB) are located on either side of the LB and cover the region 0.8 < |$\eta$| < 1.7. The Barrels are divided into two partitions for operational purposes. Each partition is comprised of 64 wedge-shaped segment modules. Light from the side of each cell is collected by wavelength shifting fibers, and read out by two PhotoMultiplier Tubes (PMTs) positioned in insertable “drawers” within the back girders of each module. Several detector components will be replaced (e.g. front-end electronics of the calorimeters) and major changes on the trigger system which is required to cope with the new luminosity requirements~\cite{ref1,ref2,ref3,ref4}.\footnote[1] {The ATLAS experiment makes use of a right-handed coordinate system with its origin located at the nominal interaction point (IP) in the centre of the detector and the z-axis along the beam pipe. The x-axis points from the IP to the centre of the LHC ring, and the y-axis points upwards. Cylindrical coordinates (r,$\phi$) are used in the transverse plane, where $\phi$ is defined as the azimuthal angle around the z-axis. The pseudorapidity is defined in terms of the polar angle $\theta$ as $\eta = -ln \tan{(\theta/2)}$.}\\
The Tile Calorimeter front-end electronics (placed inside drawers) are powered by 10V DC prepared in the finger adjacent
to the superdrawer 
inside the so called finger Low Voltage Power Supply (fLVPS). Each fLVPS also referred to as an LVBOX, contains eight DC/DC converters also called Bricks and are transforming 200V DC input into 10V DC output. The
LVBOX is water-cooled, the Bricks are affixed on both sides of the cooling plate which sink the heat inside the LVBOX. The LVBOX contains an ELMB-MB holding a ELMB chip, a Fuse Board distributing 200V DC coming from ATLAS technical cavern USA15 to the Bricks and internal cable set connecting elements inside the LVBOX with outside elements via a 72 pin Harting connector.\par
\noindent The ELMB incorporates a CAN Bus protocol for communication that allows the monitoring of\,behavioural parameters of DC/DC converters such as Input Current (I$_{in}$), Output Current (I$_{out}$), Input Voltage (V$_{in}$), Output Voltage (V$_{out}$), and two onboard temperatures (TMEAS2) and (TMEAS3). The Auxiliary boards (AUX-Boards) are placed at USA15 providing control of Bricks inside the LVBOX with a possibility of switching each Brick individually on and off.
All produced Bricks inside the LVBOX will provide the same output voltage which is the main difference in the LVPS Phase-II upgrade to the previous iteration of Bricks.
\noindent The Phase-II upgrade program is set to improve the TileCal performances according to the HL-LHC requirements. In order to cope with the increased luminosity, an upgraded LVPS Brick has been re-designed, aiming to power the on detector electronics for the increased trigger rates and high-performance data acquisition \cite{ref7, ref8}.
  

\label{sec:intro}
\begin{figure}[t]
\includegraphics[width=15cm]{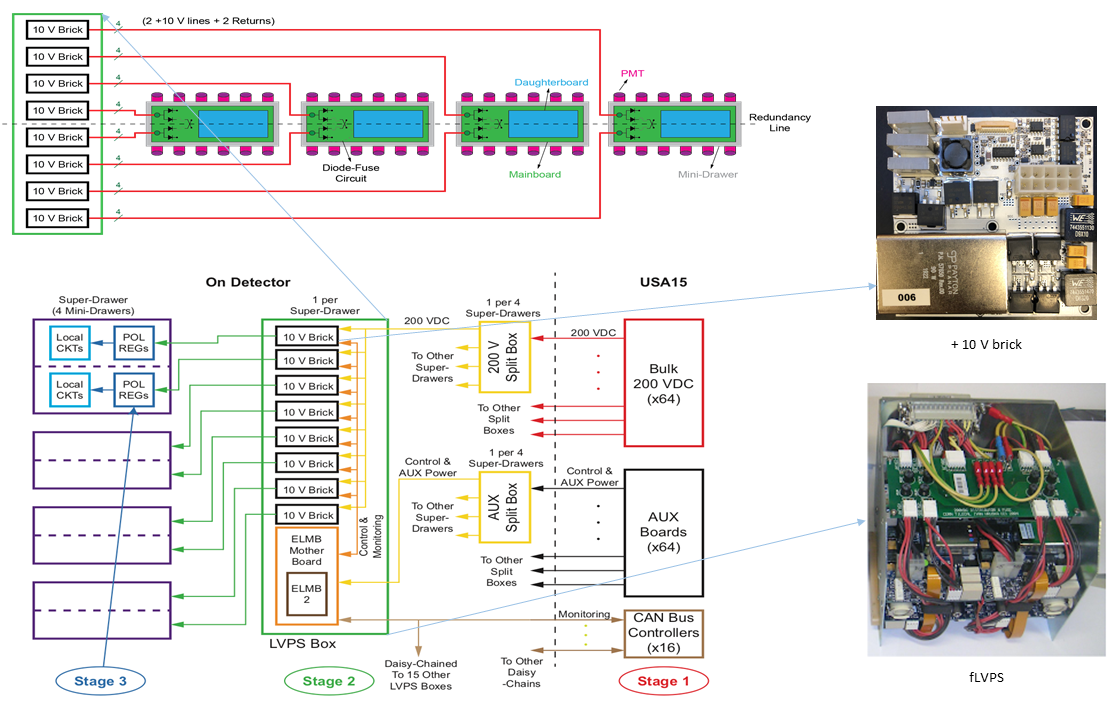}
\centering
\caption{Full diagram and schematic of the LV System which shows the Block diagram of the power system as well as the schematic of the 3-stage low voltage power distribution system. An image of the finger Low Voltage Power Supply and the upgraded LVPS Brick.}
\label{fig2}
\end{figure}
\section{Low Voltage power distribution system and test bench usage}
There are three power stages for the low voltage power distribution system (Figure~\ref{fig2}), the first step in the low voltage distribution system is produced by the bulk 200V DC commercial power supplies~\cite{ref5, ref6} located at USA15. The second step is the production of 10V DC by the upgrade Bricks inside the fLVPS as described above. The third step is the conversion of incoming 10V DC by Point of Load (POL) converters into the required voltages inside the front-end electronics~\cite{ref7}. The main challenge is the Bricks operating within TileCal have to cope with high radiation levels, to be immersed in the Barrel Toroid’s magnetic field~\cite{ref8} and have very little space available. Despite these constraints the Bricks can be built with components that are qualified to handle the radiation tolerance with a compact fit able to operate in a magnetic field as well a low-noise threshold for the HL-LHC.
In the present system, the eight Bricks that reside in the fLVPS each provide power for specific circuitry in the drawer. They all use the same basic design but are configured for the circuitry that they service, resulting in eight different types. In the new
design, all eight Bricks have the same specifications and performance requirements~\cite{ref9}.\\
\noindent Significant effort has gone into the design and testing of the current power supplies for both performance and radiation tolerance. The design changes for the upgraded prototypes have undergone various testing to retain this performance.  Full characterization of the final design will be performed before production, which will also entail using the initial test bench to facilitate the inspection aspect of quality management of the Brick. Dedicated test benches have been constructed to test any upgraded Bricks to be used for the HL-LHC. Expected features of the test bench include parametric measurements and monitoring as well as functional tests (input current, LT DC/DC convertor chip functionality and feedback loop parameters), protection circuits triggers, such as Over-Voltage, Over-Current, Over-Temperature (OVP, OCP, and OTP). This allows for quick testing to evaluate the Bricks performance metrics.
\subsection{LVPS Brick design overview}
\begin{figure}[t]
\includegraphics[width=15cm]{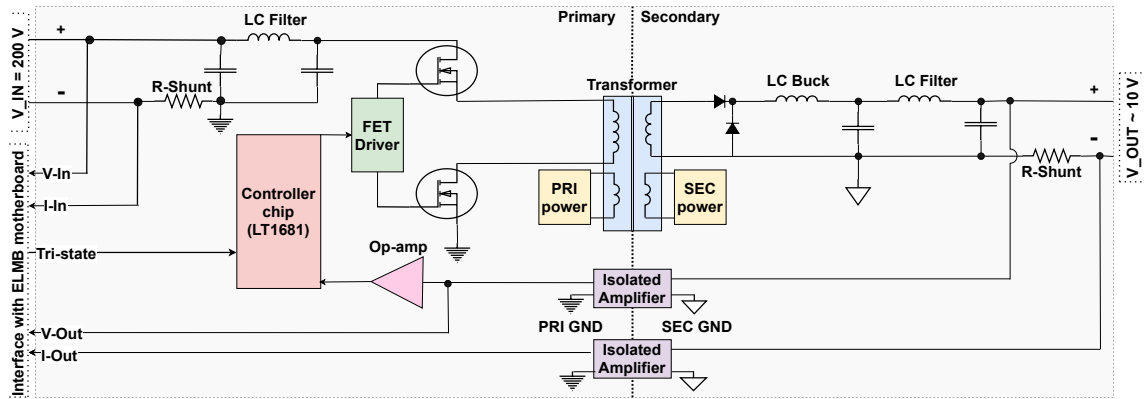}
\centering
\caption{Functional block diagram of an LVPS Brick.}
\label{fig3}
\end{figure}
The present Bricks are a refined custom design that has evolved over several iterations in recent years. All upgraded Bricks share common design requirements, which makes the testing a lot simpler for final production. The requirements of the Bricks include a higher radiation tolerance for the HL-LHC, a compact packaging fit into the foot print available for the fLVPS, being able to operate in a magnetic field together with a overall efficiency of over 80~\%. Additionally these Bricks are required to operate at a nominal output power of 230 W when 2.3 A nominal load is applied with an output voltage of 10V DC. The Brick block diagram of the circuit is shown in Figure~\ref{fig3}. The heart of the design is the LT1681 controller chip~\cite{ref11}, a dual transistor synchronous forward controller able to produce the drive pulse at the frequency of 300 KHz with an output duty factor that can vary from a few percent up to a maximum of 45\%. The pulse width is controlled by a feedback circuit based on the values of the output voltage of the Brick. This input permits us to ensure continuous-mode operation at the nominal voltages and currents. The signal from the LT1681 is sent to the Field Effect Transistor (FET) drivers. These are transistor drivers that have sufficient current and voltage drive to drive the high-side. Switching power supplies are objects able to produce DC signals with reduced voltages to the input signal while maintaining the power dissipation very low.\\ 
\begin{figure}[t]
\includegraphics[width=12.5cm]{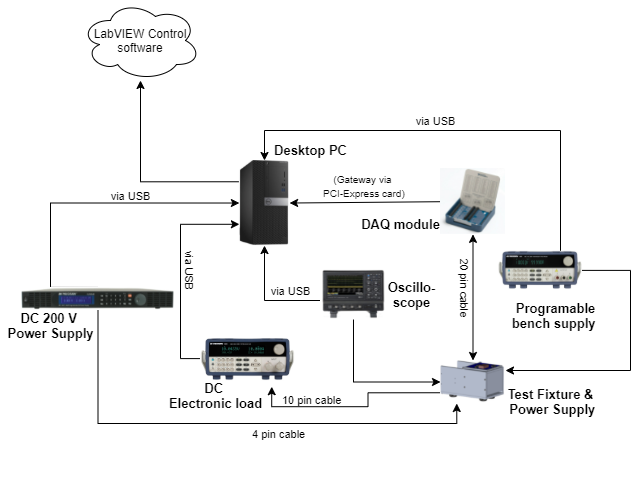}
\centering
\caption{Functional test bench block diagram.}
\label{fig4}
\end{figure}
\noindent The high-side and low-side transistors simultaneously turn on and conduct for the duration that the output clock is in the high state, and both are in the off state when the clock is low. When the MOSFETs conduct, current flows through the primary windings of the transformer, which transfers energy to the secondary windings. A buck converter is implemented on the secondary side of the transformer and converts the signal to a constant voltage for the output. The output side also contains an additional combination of inductors (L) and capacitors (C), LC stage for noise filtering. Voltage feedback, for controlling the output voltage, is provided by the optoisolators. This component processes the electrical signal converting it into a digital signal, at the input stage, and transmitting it through a pulsed light signal to the output where it is converted back to an electrical analog signal. The signal is transmitted maintaining the input and output electrically insulated from each other. The design also incorporates two shunt resistors for measuring the output current, the voltage for which is also fed back using an optoisolator. The value of the output voltage is controlled by a reference voltage that comes from the central controller in the LVPS box, and the ELMB. The Brick has three types of protection circuits built-in as part of the design, Over Voltage Protection (OVP) and over current protection (OCP) are on the primary side and are configured
specifically for the Brick. The third one is over temperature protection, which monitors the temperatures of the low-side transistor on the primary side. This circuit is integrated in the LT1681 chip design. When one protection circuit is triggered an ‘off’ signal is sent to the LT1681 which stops the Brick immediately. Several monitor circuits are implemented on the Brick. These circuits send the information to the ELMB that in case of need can command a shutdown. In particular, the ELMB monitors two temperatures on the primary side of the Brick both close to the input of over temperature protection of the LT1681 and the output values. The upgrade Brick will have a single Brick on/off control through a TRI-STATE line. The TRI-STATE control lines are driven by the power supplies where the start voltage level of the Brick ranges at 15$\pm 2$V DC and the high impedance signal enables the LM9074M regulator on the LVPS Brick. The disable state signal disables the LM9074M regulator on the LVPS Brick and switches off the Brick when switched to ground.
\section{LVPS Brick test bench system hardware architecture}
\begin{figure}[t]
\includegraphics[width=15cm]{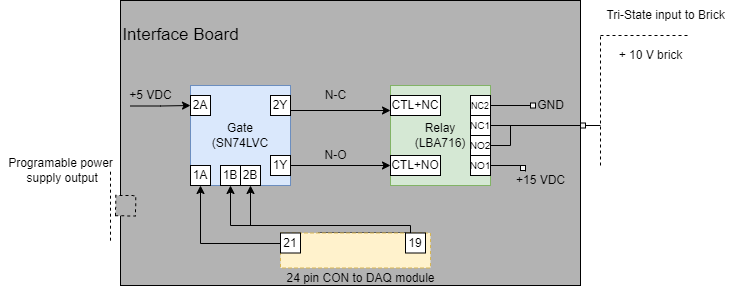}
\centering
\caption{Functional block diagram of the interface board.  A logic AND gate
(SN74LVC) and a dual solid-state relay (LBA716) are used to switch +15 VDC to the TRI-STATE
input of a Brick. A TRI-STATE represents three voltage levels to start Bricks from an Embedded
Local Monitor Board (ELMB), only an external 0 VDC and +15 VDC signal is needed to
start and stop a Brick.}
\label{fig5}
\end{figure}
The test bench was developed for the purpose of evaluating the functionality of the LVPS Bricks following their production~\cite{ref11}. The test bench provides a set of independent  parametric and functional tests (input current, LT DC/DC converter chip functionality, and feedback loop parameters) tests to be performed by personnel and would generate reports that can be stored in a database for
future analysis. The test bench consists of a test fixture providing support for both the LVPS Brick and interface card that facilitates the monitoring and control of the power supplies. The test bench is fully automated by means of a LabVIEW control program~\cite{ref12}. The test of one LVPS Brick requires only 4 minutes to go through the sequence of automated tests and guarantees full functional testing. The control software is read out and controlled by different commercial instruments. The central element of the system is a data acquisition (DAQ)~\cite{ref13} module from National Instruments which serves as the primary method for acquiring data from the Brick under test. The DAQ module communicates with the computer via a PCI-express card~\cite{ref14}, which acts as a gateway for the PC to the Data Acquisition module. In addition to the Data Acquisition module, the test stand has several other instruments, including an electronic load meter which provides a programmable load for testing the 
performance of the upgraded Bricks. A programmable power supply is used for providing +5V for peripheral 
circuitry of the interface card, including the bias voltage for the temperature sensors as well as TRI-STATE voltages to start. A programmable high voltage power supply is used to provide the 200V needed to power the Brick 
under test. Lastly, a digital oscilloscope is used for reading out waveforms, clock and the dynamics of the output voltage. These instruments are controlled and read out using either a USB/RS232 or GPIB bus. 
The basic system topology is shown in Figure~\ref{fig4}. 
\subsection{Interface card and Data Acquisition module}
\begin{figure}[t]
\includegraphics[width=13.5cm]{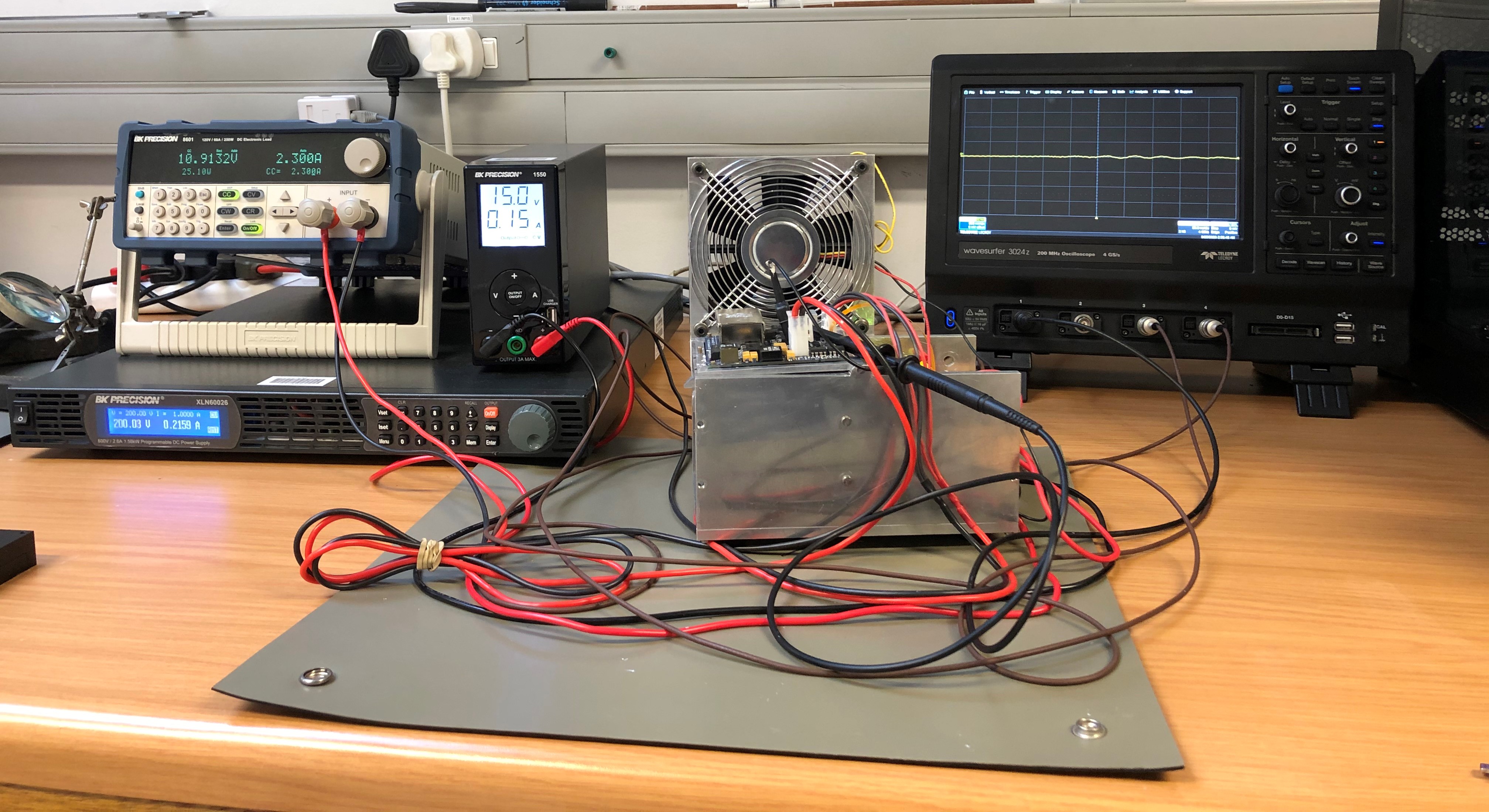}
\centering
\caption{Automated test bench for an LVPS Brick.}
\label{fig6}
\end{figure}
The interface card (see Figure~\ref{fig5}) acts as an interface between the Brick, a programmable bench top power supply and the DAQ module. The interface circuitry has been updated to include TRI-STATE signal functionality which in normal conditions would be coming from the auxiliary board through the ELMB-MB to start the Brick, keep it on, or shut it down. This allows for the Brick to be powered on or off via the TRI-STATE signal line in an fLVPS. The second allows for the on/off control of the input 200$\,$V DC provided to the Brick by the DC power supply.  A metal case that acts as Brick support and provides the interface to the computer and the ground connections. The DAQ module SCB-68A is used as part of the input/output (I/O) current, voltage and temperature attenuation measurements connected to the computer using a Peripheral Component Interconnect (PCI) interface. The shielded I/O connector block has 68 screw terminals for easy signal connection to the interface card. The data acquisition module digitizes eight differential channels, as well as sending and receiving digital bits that are used for control purposes.  The interface board data acquisition involves reading four differential
inputs; input current (I$_{in}$), input voltage (V$_{in}$), output current (I$_{out}$), output current (I$_{out}$) and two temperature measurements (TMEAS2 and TMEAS3) from the Brick. The signals are transmitted from the Brick to the interface card via the DAQ and PC.

\section{Automated testing program}
A real-time program was modified to read the status and the internal test results of all the individual Bricks under test. The test program can also evaluate the data read for each test, and applies cuts to determine if the measurements meet the acceptance criteria. The program is written graphically in a National Instruments LabVIEW framework, together with drivers provided by the commercial instruments as well as the routines for communication with the DAQ module. The program also checks the functioning of the LVPS Bricks (current, voltages and temperature). The program evaluates the Brick's operation and checks the data collected by commercial equipment, which are used to perform the procedure. It also saves the data for each Brick test.
The test bench is based on a computer controlling and reading out several commercial equipment which performs the tests. Eleven separate tests in total complete the procedure, each communicating with several instrumentation devices. The full composition of the test stand can be seen in Figure~\ref{fig6}.
\begin{figure} [t]
    \centering
    {{\includegraphics[width=7.0cm]{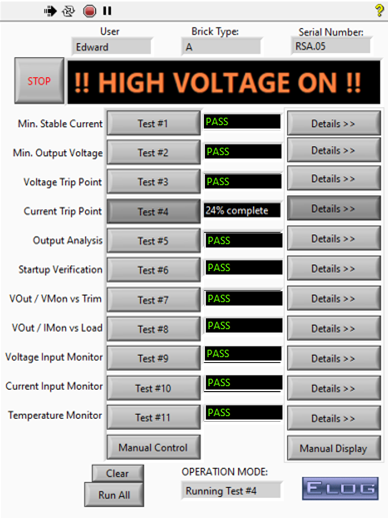}}}
    \qquad
    {{\includegraphics[width=7.0cm]{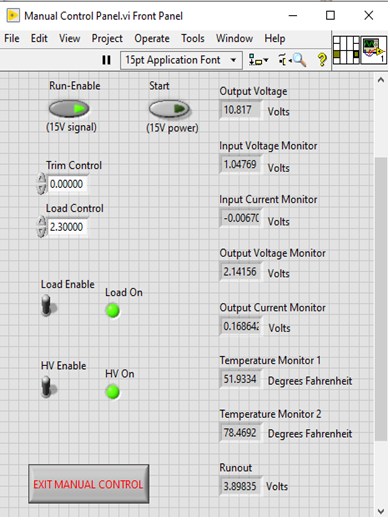}}}
    \caption{Test panel to perform various tests (left) and the manual control feature on a Graphical User Interface (right).}
    \label{fig7}
\end{figure}
\newline 
\subsection{Description and functionality of tests} In the initial and final tests~\cite{ref6,ref7} we check that the main functions and parameters (shown in Table~\ref{tab1}) of the Bricks are correct. The required parameter range corresponds to 3$\sigma$ around the nominal value for all tests except the over voltage and over current protections. For these tests the requirement is set by quality assurance procedures. In the following a brief functionality of each test is given however the goals of these tests can be summarized as follows: 
Ensure that the Brick responds correctly to regular start-up and shut down procedures commands and that it respects the load parameter (tests 1, 6); check the voltages and the currents at the input and output stages, verify the functioning of the relative monitor and ensure that the Brick responds correctly to the external reference changes (tests 2, 7, 8, 9, 10); check the output and the feedback sent to the LT1681, to ensure the stability of the
output (test 5); ensure the correct behavior of the over voltage and over current protection circuits by forcing the Brick in the condition to trigger the circuits (tests 3, 4) verify the functioning of the temperature monitors (test 11). The following are the test purposes: \par
\begin{table} [t]
\center
\begin{tabular}{lll} 
    \hline\hline
    Parameter & Minimum & Maximum \\
    \hline
      Frequency Standard Deviation (Hz) & 0 & 1000 \\
      Frequency Max (Hz)  & 290000 & 350000 \\
      Frequency Min (Hz) & 250000 & 310000 \\
      Minimum Stable Load (A) & 0 & 2.3 \\
      Minimum Output Voltage (V)  & 9.8 & 10.9 \\
      Over Voltage Protection Trip Point (V) & 11.5 & 12 \\
      Over Current Protection Trip Point (A) & 10.25 & 10.75 \\
      Duty Cycle Standard Deviation (\%) & 0 & 0.1 \\
      Clock Duty Cycle Average (\%) & 0 & 45\\
      Clock Duty Cycle Standard Deviation (\%) & 0 & 0.15\\
      Maximum Start-up Delay (s) & 0.2 & 0.08\\
      \hline \hline
      \centering
      \end{tabular}\\
      \caption{Testing station test parameter bounds. These parameters are used to restrict the response of the output to a change in measurement when evaluating the LVPS Brick under test.}\label{tab1}
\center
\end{table} 
\begin{enumerate}
\item  The minimum stable load test is designed to determine the minimum load that the power supply is behaving correctly. In this condition, the output from LT1681 is read with the oscilloscope. The load is then decreased until missing clock cycles are registered, the load value is recorded. Permitted values for the test depend on the Brick
type.
\item  Minimum Output Voltage test measures the minimum output voltage. The
lower limit for each Brick is shown in Table \ref{tab1}. These values are within a few percent of
the nominal output\,values.
\item The voltage trip point test checks the over voltage protection circuitry of the Brick. The test simulates an over voltage (for the output voltage) triggering the protection circuit that trips the Brick. The condition of over voltage is reached decreasing the external reference value. The test records the voltage when the Brick
trips. The protection circuit is set to shutdown the Brick at 10-20~\% over the nominal operating voltage.
\item The current trip point test simulates an over current event which would cause the protection circuit to trip the Brick. The test simulates an over current event (for the output current) triggering the protection circuit that trips the Brick. The condition of over current is reached decreasing the load value. The test records the current when the Brick trips. The protection circuit is set to shutdown the Brick at 25-50~\% over nominal operating current.
\item The output stability test is used to check the stability of the output and the signal from the Brick. The test also measures the frequency and duty cycle of the driving pulse. The output voltage, the frequency and the duty cycle of the driving pulse are measured over several iterations. The maximum output voltage RMS is then required to be within the values listed in table~\ref{tab1} The driving pulse frequency must be within 10~KHz range of the nominal frequency of 300~KHz. The duty
cycle for the 10V Brick generally accepted RMS value is 0.1~\% for all Bricks. In table \ref{tab1} the desired duty cycle factors
for each Brick are shown. 
\item The startup verification test measures the delay that the Brick takes to start-up. It is typically measured when the voltage reaches 90~\%. For this measurement the oscilloscope probes are connected to the output and to the input of the Brick to register the pulse, the delay is measured when the voltage amplitude reaches 90\%. Start-up delay acceptable values are shown in Table~\ref{tab1}.
\item The reference voltage is also measured at regular intervals to determine the response of the
output of the external reference and checks the correct functioning
of the output voltage monitor. In the test, we measure the slope and offset of the output voltage (V$_{out}$) versus the reference, then we take the same measurements for V$_{out}$ monitor versus offset. The output has a well defined slope in the process which allows us to check that the output monitor sees the
changing in the output and to define the proportional coefficient between the monitor and V$_{out}$. 
\item The output current versus load test checks the correct working of the monitor's circuit. The Brick is started at the minimum output voltage and the load is increased to 80\% of the trip point. Output of I$_{out}$monitor is taken at regular intervals of the load. The test measures the slope and offset of the plot.
\item The calibration test is performed to make sure that the voltage input monitor is set to the correct value. The output voltage is set to the nominal value and the test measures the value of the input voltage monitor. The correct value is 1~V for all Bricks with a tolerance of ±0.05~V.
\item Current Input Monitor This test checks the correct calibration of the current input monitor. V$_{out}$ is set to the nominal value and the test measures the value of the input current (I$_{in}$) monitor.
\item The temperature monitor test is used to check the temperature of the monitor located near the input of LT1681 temperature protection. The Brick is cooled using a fan during the test therefore the measured values do not correspond to the operation values.
\end{enumerate} 
\noindent
The manual control feature has been included along with an Electronic LOG book feature which is a widely used service on many CERN experiments. A LabVIEW-based program has been implemented into both the test bench to automate new ELOG entries containing test parameters immediately after a LVPS Brick test is completed. The Test bench logs 50 parameters of each Brick in a single ELOG entry. The ELOG ‘client’ is run locally on the test bench, and pipes parameters to the ELOG ‘daemon’ on the local PC.

\begin{figure}[t]
\includegraphics[width=13cm]{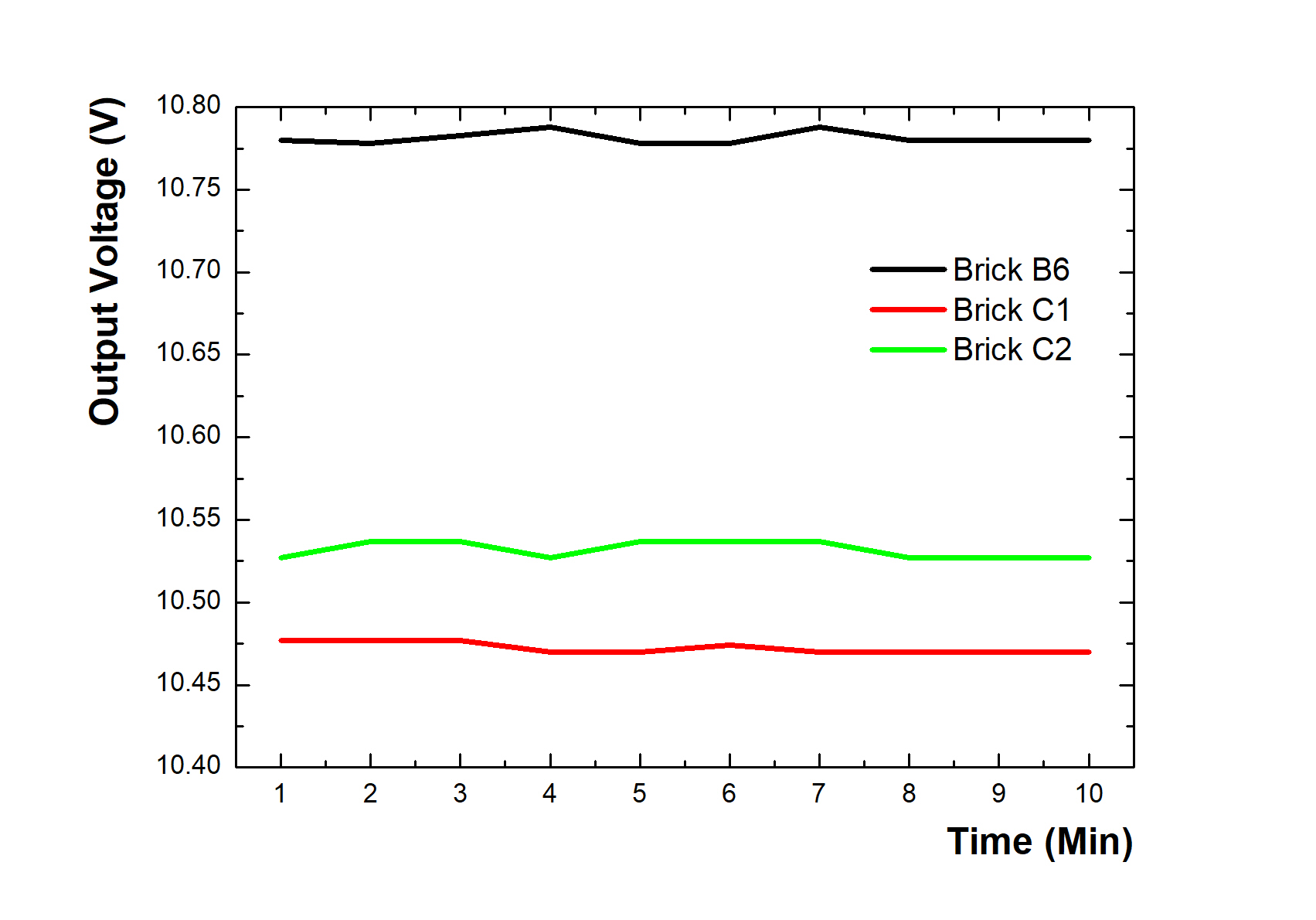}
\centering
\caption{The test monitors the output voltage of the LVPS Brick. Permitted values for all the Bricks should be within the 3$\sigma$ range. The test was recorded for 10 minutes.}
\label{fig8}
\end{figure}

\section{Monitoring and control results}
\begin{figure}[t]
\includegraphics[width=13cm]{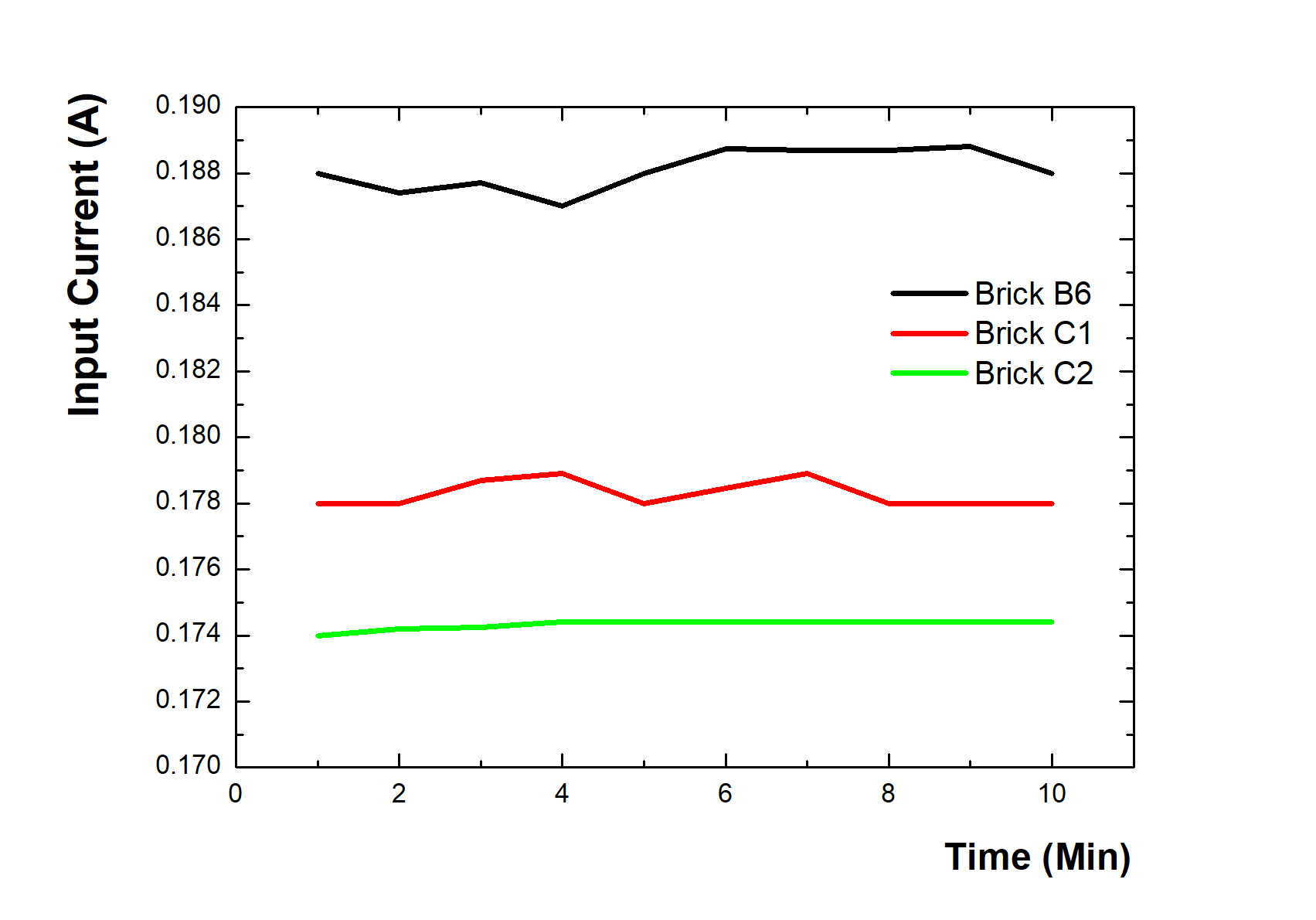}
\centering
\caption{The test monitors the input current of the LVPS Brick. Permitted values for all the Bricks should be within the 3$\sigma$ range. The test was recorded for 10 minutes.}
\label{fig9}
\end{figure}
Some notable metrics that are also measured are the clock and its jitter which are monitored to see how they affect the system's stability and to make sure that the LVPS Brick protection circuitry is operating properly. 
This test bench also verifies the protection circuitry of LVPS Brick, which protects it from over temperature, current and voltage. If the output voltage, current or the temperature of the Brick exceeds the allowed range it can damage the Brick itself, as well as to the components it is powering. If the temperature or the output voltage of the Brick exceeds the tolerance limit it can damage the components of the system and reduce the life expectancy of the Brick.
The test process is initiated by the user with the required test. After the test is completed, the program then automatically configures all the necessary parameters and displays a binary pass or fail alongside the respective test as shown in Figure \ref{fig6}.
The program can also export the results as a text file or if required in a form of a graphical representation. 
For the completion of a test sequence, all tests need to be successfully completed without any error messages being displayed. 
The pass status all of its respective tests must have yielded results within the allowed parameter distribution, that is, all eleven tests within the sequence must have received a pass.\par
\begin{figure} [t]
    \centering
    {{\includegraphics[width=7.1cm]{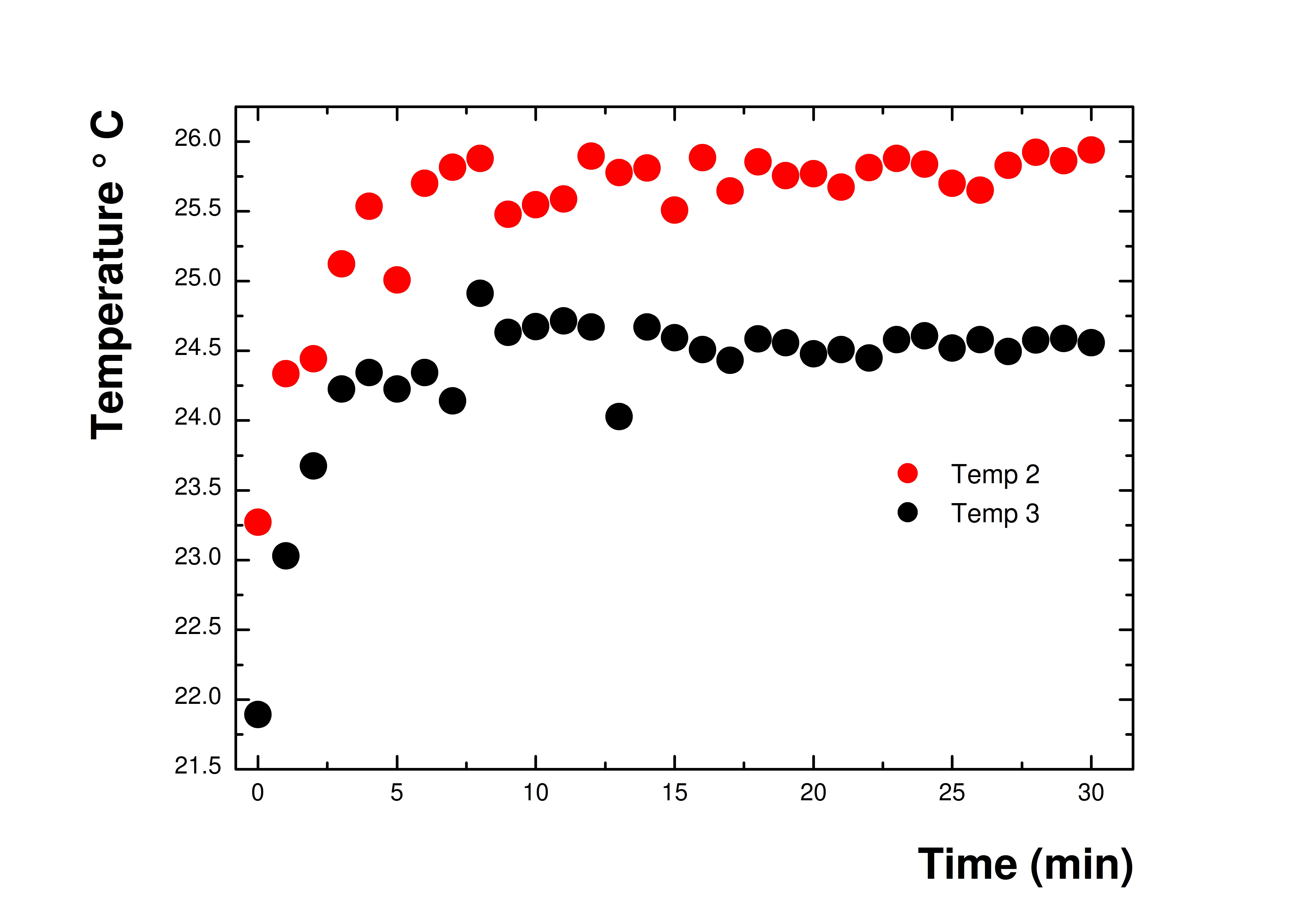}}}
    \qquad
   {{\includegraphics[width=7.1cm]{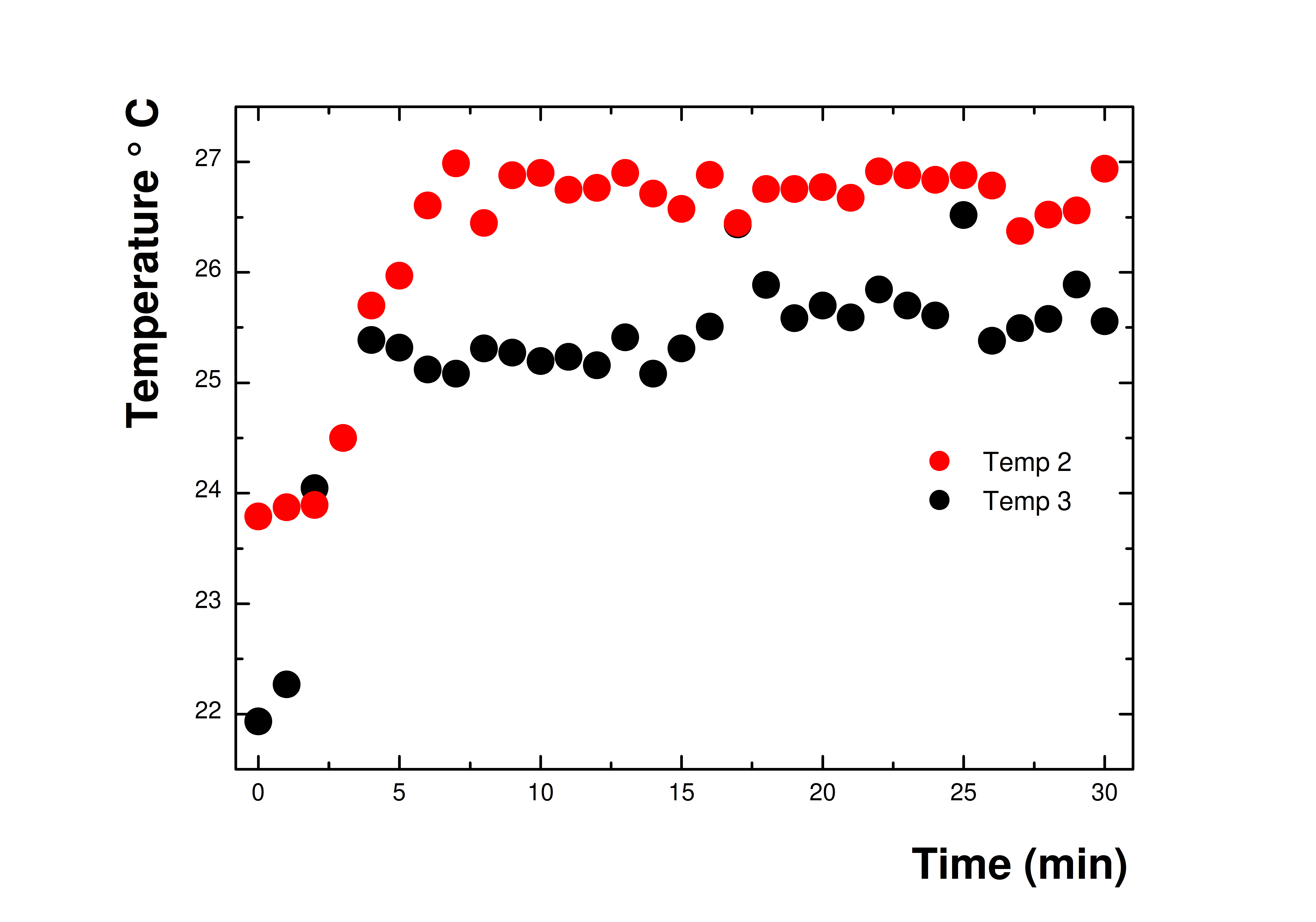}}}
    \caption{Temperature measurements were taken for continuously for 30 minutes. Brick B6 (left) and Brick C1  (right) single-ended temperature measurements show relatively low temperatures whilst being cooled at 20\,$^{\circ}$C as would be under normal operation of ATLAS.}
    \label{fig10}
\end{figure}

 Another significant metric we are measuring is the signal feedback where the test checks the correct functioning of the monitor circuit of output voltage and input current (see results in Figure~\ref{fig8} and \ref{fig9}, respectively) to be distributed at the front end electronics of the TileCal. The Brick is started at a nominal load of 2.3 A. The manual control feature is a new addition, which allows the Brick to be monitored at any period interval.
Permitted values for all the Bricks should be within the 3$\sigma$ range.  For values outside the specified values, these are considered as a failure of a Brick. The expected voltage of the power supply should not exceed 12.5 V, which can be observed from Figure~\ref{fig8} where both the upper and lower bounds of the tested Brick are with the required specifications. The same can be observed for the input current of the Brick which should not exceed a value of 0.220\,A as seen in Figure~\ref{fig9}. The two single-ended temperature measurements (TMEAS2 and TMEAS3) for the prototype Bricks under test can be observed in Figure~\ref{fig10}.
\section{Conclusion}
The ATLAS TileCal collaboration is currently undertaking extensive preparations for the Phase-II upgrade in order to allow the full exploitation of the upcoming HL-LHC physics opportunities. Several Bricks have been manufactured by Wits University and have passed all tests of the quality assurance test station. The testing bench has been fully developed and commissioned alongside the Phase-II upgrade Brick to this end. The test benches will ensure the high quality and reliability of the 2048 Phase-II upgrade Bricks once installed on-detector. Each LVPS Brick will undergo a long list of tests to ensure the correct functioning of the Bricks. Any departure from the stated performance metrics will result in the Brick in question being repaired and retested in order to avoid the eventuality of a TileCal modules on-detector electronics being offline. These Bricks were
also exposed to the Super Proton Synchrotron accelerator
at CERN for the test beams of 2018.
\vspace{0.5cm}
\acknowledgments
The authors are grateful for the support from the South African Department of Science and Innovation through the SA-CERN program and the National Research Foundation for various forms of support.
\vspace{1em}\\
\tiny $\copyright$ Copyright owned by the
author(s) under the terms of the Creative Commons\vspace{1em}\\ Attribution-Non Commercial-No Derivatives 4.0 International License (CC BY-NC-ND 4.0).

\end{document}